\documentclass[twocolumn,superscriptaddress,prl]{revtex4-2}
\usepackage{amsmath}
\usepackage{graphicx}
\usepackage{xcolor} %color stuff
\usepackage[normalem]{ulem} %strike through text

\newcommand{\stkout}[1]{\ifmmode\text{\sout{\ensuremath{#1}}}\else\sout{#1}\fi}
\newcommand{\md}{\mathrm d}

\newcommand{\mcQ}{\mathcal Q} % Heat
\newcommand{\mcV}{\mathcal V}

\begin{document}

\title{Mechanical loss in amorphous solids: spatial correlations, interacting transitions, and annealed thermodynamic pathways}

\author{Steven Blaber}
\email{steven.blaber@ubc.ca}
\affiliation{Dept.~of Physics and Astronomy and Stewart Blusson Quantum Matter Institute, University of British Columbia, Vancouver, British Columbia V6T 1Z1, Canada}
\author{Jörg Rottler}
\email{jrottler@physics.ubc.ca}
\affiliation{Dept.~of Physics and Astronomy and Stewart Blusson Quantum Matter Institute, University of British Columbia, Vancouver, British Columbia V6T 1Z1, Canada}

\begin{abstract}
The disordered and defect-rich structure of amorphous solids forms heterogeneous, high-dimensional energy landscapes. Such an energy landscape can be described by a discrete-state network of transitions between stable energy minima. Under low-frequency mechanical oscillations, defect-mediated, thermally activated transitions provide a microscopic mechanism for mechanical dissipation that are the dominant cause of mechanical loss in the mirror coatings of ground based gravitational waves detectors.  Using molecular simulations, we find spatially correlated and strongly interacting transitions that require a general network description instead of a superposition of independent two-level systems as traditionally assumed. An annealing study combined with an analysis of dominant relaxation paths in the energy landscape reveals novel mechanisms for reducing room temperature mechanical loss.
\end{abstract}
\date{\today}

\maketitle

\section{Introduction}
Internal friction and energy dissipation in amorphous solids arise from localized, atomic-scale processes that convert elastic or mechanical energy into heat. 
Materials with low mechanical loss are of great importance in numerous applications. A prime example is the reduction of thermal noise in the amorphous metal oxide mirror coatings of Laser Interferometer Gravitational Wave detectors (GWD)~\cite{steinlechner2018,vajente2021low}, but there are also other applications in cavity optomechanics and sensors~\cite{aspelmeyer2014}. The precise mechanism that causes dissipation in disorderd solids varies greatly with the frequency band of interest. While nonaffine coupling of strain to vibrational degrees of freedom dominates at high frequencies~\cite{milkus2017,zaccone2023, damart2017}, thermally activated transitions of clusters of atoms between energy minima in the potential energy landscape are believed to provide the major contribution at low frequencies~\cite{damart2018}. When these transitions occur far apart and in isolation from each other, they can be thought of as two-level-systems (TLS). The linear response of such an ensemble of non-interacting TLS relaxing back to equilibrium after excitation by oscillatory external fields was computed over 50 years ago and forms the basis of the celebrated TLS model~\cite{phillips1987}.

The TLS model has provided the foundation for the interpretation and analysis of numerous experimental studies of mechanical loss.  The current GWD mirror coatings use amorphous TiO$_{2}$:Ta$_{2}$O$_{5}$ and SiO$_{2}$ as high and low refractive index materials, respectively. Other relevant amorphous materials include \textit{a-}Si, a very low loss material~\cite{liu2014hydrogen} that is considered for optical stacks for GWD mirror coatings~\cite{steinlechner2015thermal,molina2025revealing}, and Ti:GeO$_{2}$, a leading candidate for next generation GWDs due to its low loss and adequate refractive index~\cite{vajente2021low}. For instance, the temperature dependence of mechanical loss in Ta$_{2}$O$_{5}$, Ti-doped Ta$_{2}$O$_{5}$, and SiO$_{2}$ coatings has been analyzed in terms of thermally activated TLS and distributions of activation barriers~\cite{martin2009comparison,martin2010effect,martin2014low}. Related experimental studies have used the TLS picture to interpret cryogenic loss peaks and their suppression through changes in composition, layering, or hydrogenation in candidate coating materials such as composite SiO$_{2}$, GeO$_{2}$, Ti$_{2}$, and \textit{a-}Si~\cite{kuo2019low,khadka2023,molina2021,molina2023}. These studies provide an important phenomenological connection between measured loss spectra and microscopic structural excitations.

In parallel, atomistic simulations have been performed to identify structural defects and compute mechanical loss of SiO$_{2}$ \cite{damart2018}, undoped \cite{puosi2019silico, trinastic2016} and doped Ta$_{2}$O$_{5}$ ~\cite{trinastic2016,prasai2019high,jiang2021analysis} within the TLS model. Amorphous silicon has also been studied in detail, with atomic scale transition and defect motifs identified that cause mechanical loss in the TLS model~\cite{levesque2022,girard2025}. The structural characteristics of defects and their connection to mechanical loss has helped inform our understanding of mechanical loss in Ti:GeO$_{2}$~\cite{prasai2026atomic}.

In a recent contribution, we reported results from atomistic simulations of amorphous silicon and titanium dioxide that are incompatible with the dissipation arising from isolated, noninteracting TLS~\cite{Blaber2026}. By searching for transitions in the energy landscape, we discovered that TLS form a sparse, scale free connected network whose structure is very different from that produced by many independent TLS. We then computed, in linear response, the dissipation produced by the networks, and showed that our prediction for the mechanical loss systematically generalizes the TLS model. The connectivity of the network introduces new mechanisms that can both reduce low frequency dissipation through additional low energy relaxation pathways, and increase dissipation through a broad distribution of energy minima. These arise because it is the entire ensemble of transitions between of connected inherent states with its full spectrum of relaxation times and occupation probabilities that comes in resonance with the mechanical wave, and not just a superposition of individual TLS.

The purpose of the present work is to clarify and expand our proposed connected network perspective with additional results and analysis. After briefly recapitulating the analytical development of the generalized formula for mechanical loss from connected networks in section~\ref{sect: Theory}, we provide direct evidence of interactions between TLS in Section~\ref{sect: Spatial}. Many transitions occur in close spatial proximity and influence each other through elastic and electrostatic interactions. These interactions can only be described correctly within a network description. In section~\ref{sect: Quench}, we explore the effect of quench rate, and show that the experimental trend of reduced mechanical loss with increasing relaxation~\cite{molina2023} is reproduced within the connected network model through a narrowing of the distribution of inherent structure energies. An analysis of the relaxation pathways afforded by the network provides further insight into how annealing reduces mechanical loss. 

\section{Mechanical loss of connected networks}\label{sect: Theory}

Brownian thermal noise provides the dominant loss channel in the amorphous mirror coatings of ground based GWDs~\cite{steinlechner2018,vajente2021low}. The scale of these thermal \emph{fluctuations} is quantified by the \emph{dissipation} from driven acoustic oscillations through the \emph{fluctuation-dissipation relation}~\cite{levin1998internal}. Therefore, the same structural rearrangements that are believed to be responsible for mechanical loss in amorphous solids also produce equilibrium thermal fluctuations. For any system driven out of equilibrium by acoustic vibrations, internal dissipation mechanisms result in heat $\mcQ_{\rm cycle}$ released in each oscillation cycle. Relating the released heat to the energy stored in the mechanical driving defines the inverse quality factor
\begin{align}
	Q^{-1} = \frac{1}{2\pi}\frac{\mcQ_{\rm cycle}}{{\rm energy~ stored}} \ .
	\label{eq: quality factor definition}
\end{align}
The inverse quality factor is what we refer to as \emph{mechanical loss}: a good quality material wastes only a small fraction of energy in each cycle and has low mechanical loss and inverse quality factor.

To quantify the heat dissipation and determine the mechanical loss, we use a discrete-state network representation of the energy landscape~\cite{Blaber2026}. The stable minima of the energy landscape (inherent structures) are represented by a node in the network with occupation probability $P_{i}$ and allowed transitions by an edge with transition rate $R_{ij}$ for transitions from node (state) $j$ to $i$. The dynamics therefore satisfy the \emph{master equation}~\cite{Gardiner,angelani1998connected,banerjee2012characterization}
\begin{align}
	\frac{\md \boldsymbol{P}_{t}}{\md t} = R(t)\boldsymbol{P}_{t} \ ,
	\label{eq: ME General}
\end{align}
with the transition rates expressed as Arrhenius rates in terms of the energy barriers $V_{ij}$ and energy levels $E_i$ of the inherent structures $i$ and $j$. For a system at inverse temperature $\beta \equiv (k_{\rm B} T)^{-1}$ with temperature $T$ and Boltzmann constant $k_{\rm B}$, the transition rate matrix has elements
\begin{align}
	&R_{ij} =k_{ij}e^{\beta E_{j}}\left[e^{-\beta V_{ij}}(1-\delta_{ij}) - \delta_{ij}\sum_{\ell\neq j}e^{-\beta V_{j\ell}}\right] \ .
	\label{eq: transition matrix}
\end{align}
Due to the exponential dependence, the transition rates are dominated by the barrier heights $V_{ij}$ rather than the bare transition rates $k_{ij}$~\cite{trinastic2016}. For simplicity, we assume equal bare transition rates $k_{ij} = k_{0}$ for all transitions in our numerical calculations, an assumption made in a previous study~\cite{levesque2022} and 
supported by atomistic simulations of amorphous silicon~\cite{valiquette2003}.

Since $R_{ij}$ is a transition rate matrix satisfying detailed balance, the equilibrium distribution is $\Pi_{i} = \exp(-\beta E_{i})/\sum_{j} \exp(-\beta E_j)$ and we can simplify the determination of eigenvalues and eigenvectors with the symmetric form
\begin{align}
    S_{ij} = \Pi_{i}^{-1/2}R_{ij}\Pi_{j}^{1/2} \ .
\end{align}
This symmetric transition rate matrix is Hermitian with eigenvectors $M_{ij}$ with inverse $M_{ij}^{-1} = M_{ji}$ and eigenvalues $\lambda_{i}$ corresponding to relaxation rates $\tau_{i} \equiv \lambda_{i}^{-1}$.

For a system with $N$ inherent structures, we showed in ref.~\cite{Blaber2026} that the inverse quality factor of the connected network of inherent states ~\eqref{eq: quality factor definition} is, within linear response,
\begin{align}
	Q_{\rm CN}^{-1} = \frac{\beta N\gamma_{0}^2}{4\mcV C} \boldsymbol{g}^{\top} M D(\omega\tau) M^{\top}\boldsymbol{g} \ . \label{eq: CN quality factor}
\end{align}
We have defined the probability weighted coupling with elements $g_{i} \equiv \Gamma_{i}\Pi^{1/2}_{i}$, and the diagonal spectral matrix with elements
\begin{align}
    D_{i}(\omega\tau) \equiv \frac{\omega\tau_i }{\left[1+\left(\omega \tau_i\right)^2\right]} \ .
\end{align}

\section{Spatial and network structure}\label{sect: Spatial}

Our data set consists of molecular dynamics simulations of \textit{a-}Si and \textit{a-}TiO$_{2}$ prepared via melt-quench as described in detail in Ref.~\cite{Blaber2026}. For \textit{a-}Si, 10 samples of 1000 atoms are prepared at a quenchrates of $10^{10}$, $10^{11}$, and $10^{12}$ K/s, for a total of 30 samples. For \textit{a-}TiO$_{2}$, 10 samples of $750$ atoms are prepared at a quenchrate of $10^{11}$ K/s. \textit{a-}Si is modeled using a Tersoff potential~\cite{tersoff1989}, while a Buckingham potential is used to describe \textit{a-}TiO$_2$~\cite{matsui1991molecular}. 

We find inherent structures by thermal search trajectories at $600K$ for $200$ps with a sampling frequency of $100$fs. We run $100$ search trajectories per sample for \textit{a-}Si and $200$ for \textit{a-}TiO$_{2}$. We determine candidate transitions based on changes in the minimum energy and filter them based on participation ratio and maximum atomic displacement to remove unlikely candidates. Duplicate pairs of transitions are determined and removed based on a root-mean-squared atomic displacement criterion between structures less than $10^{-4}$\AA~for \textit{a-}Si. For \textit{a-}TiO$_{2}$ we use a displacement criterion of $10^{-3}$\AA~ and an additional energy difference criterion of $10^{-3}$eV. Energy barriers between connected inherent structures (transitions) are determined by nudged elestic band (NEB) calculations and the LAMMPS code is used for all simulations~\cite{LAMMPS}.

\begin{figure}
	\includegraphics[width=\linewidth]{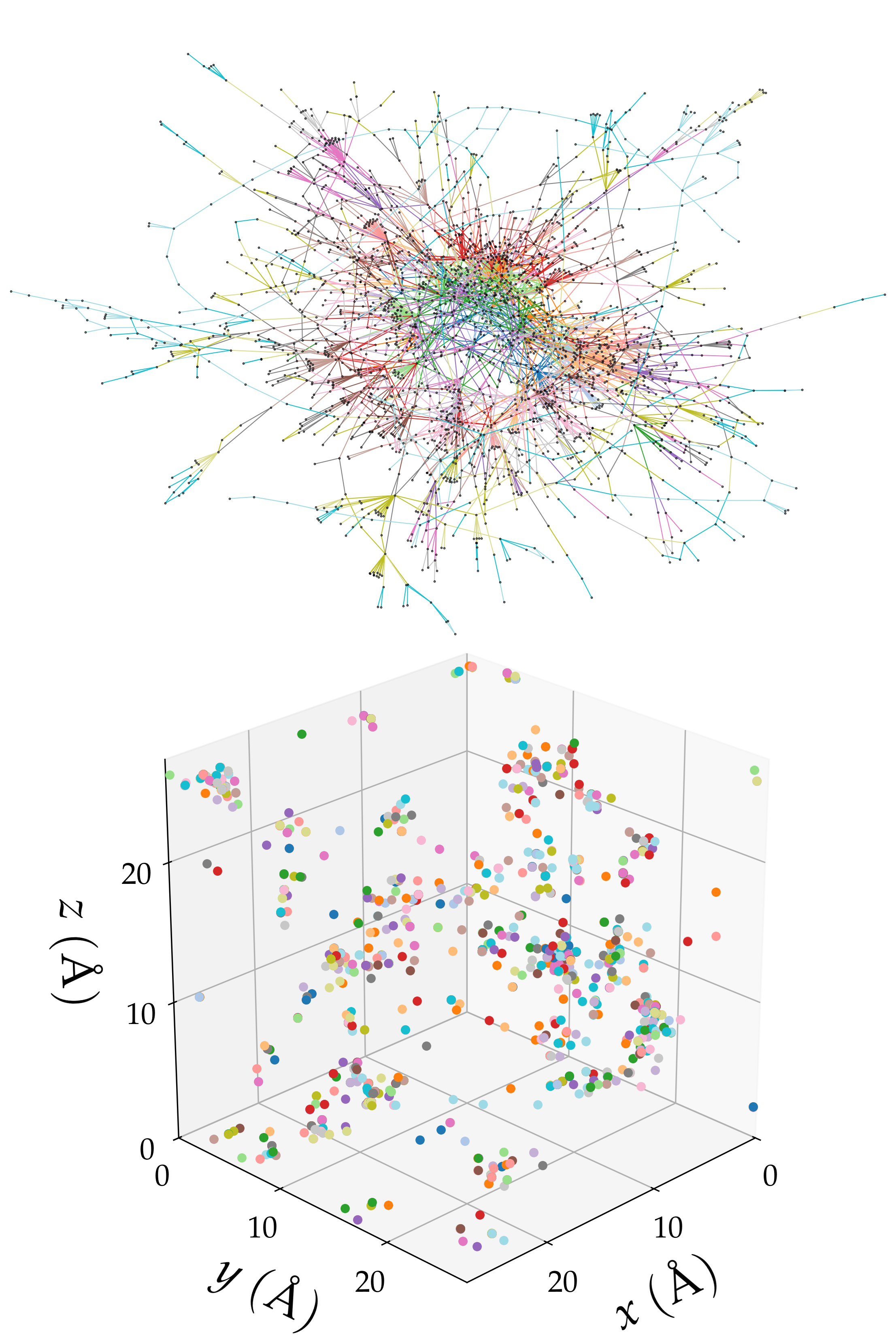}
    \caption{Top: network of transitions involved in every transition for a sample of \textit{a-}Si.  Bottom: average position of the active atoms. Colors are matched between the top and bottom sub figures, and correspond to the number of steps from the center of the network.}
	\label{fig: Full distribution}
\end{figure}

A benefit of thermal search trajectories is that they naturally and sequentially visit connected energy minima in the vicinity of the initial state. Representing each energy minimum by a point (node) and each allowed transition by a line (edge), the local energy landscape forms a network (Fig.~\ref{fig: Full distribution} top). We translate the abstract network space back to 3D real-space by computing the average position of the 10 most active atoms involved in each transition, averaged over both initial and final locations, shown in Fig.~\ref{fig: Full distribution} (bottom). 

\begin{figure}
	\includegraphics[width=\linewidth]{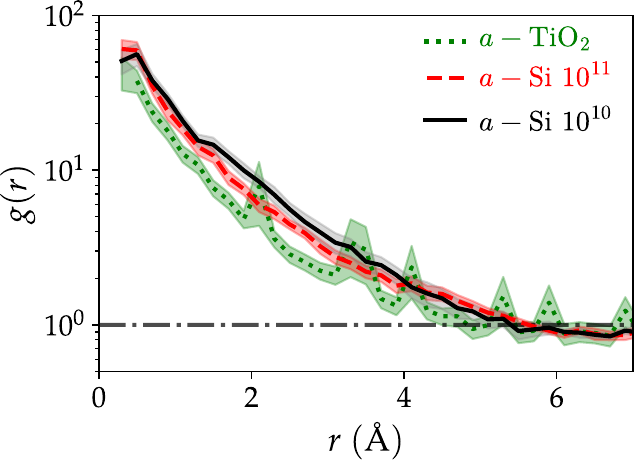}
    \caption{Radial distribution function of the mean transition locations for \textit{a-}TiO$_{2}$ (green dotted) and \textit{a-}Si ($10^{10}$K/s black solid, $10^{11}$K/s red dashed) averaged over 10 independent samples each. Shaded regions are the standard error of the mean.}
	\label{fig: RDF}
\end{figure}

To quantify the clustering of transitions, we compute the radial distribution function $g(r)$ of the mean transition locations averaged over ten samples each of \textit{a-}Si and \textit{a-}TiO$_{2}$ in Fig.~\ref{fig: RDF}. For either material, the transitions cluster within a radius of $\lesssim 4$\AA, with a peak of $\sim 60\times$ more transitions than a random (ideal gas) distribution at $\sim 1$\AA. Given the relatively high propensity for transitions to cluster at short distances, one might expect that these transitions influence each other.

\begin{figure}
	\includegraphics[width=0.8\linewidth]{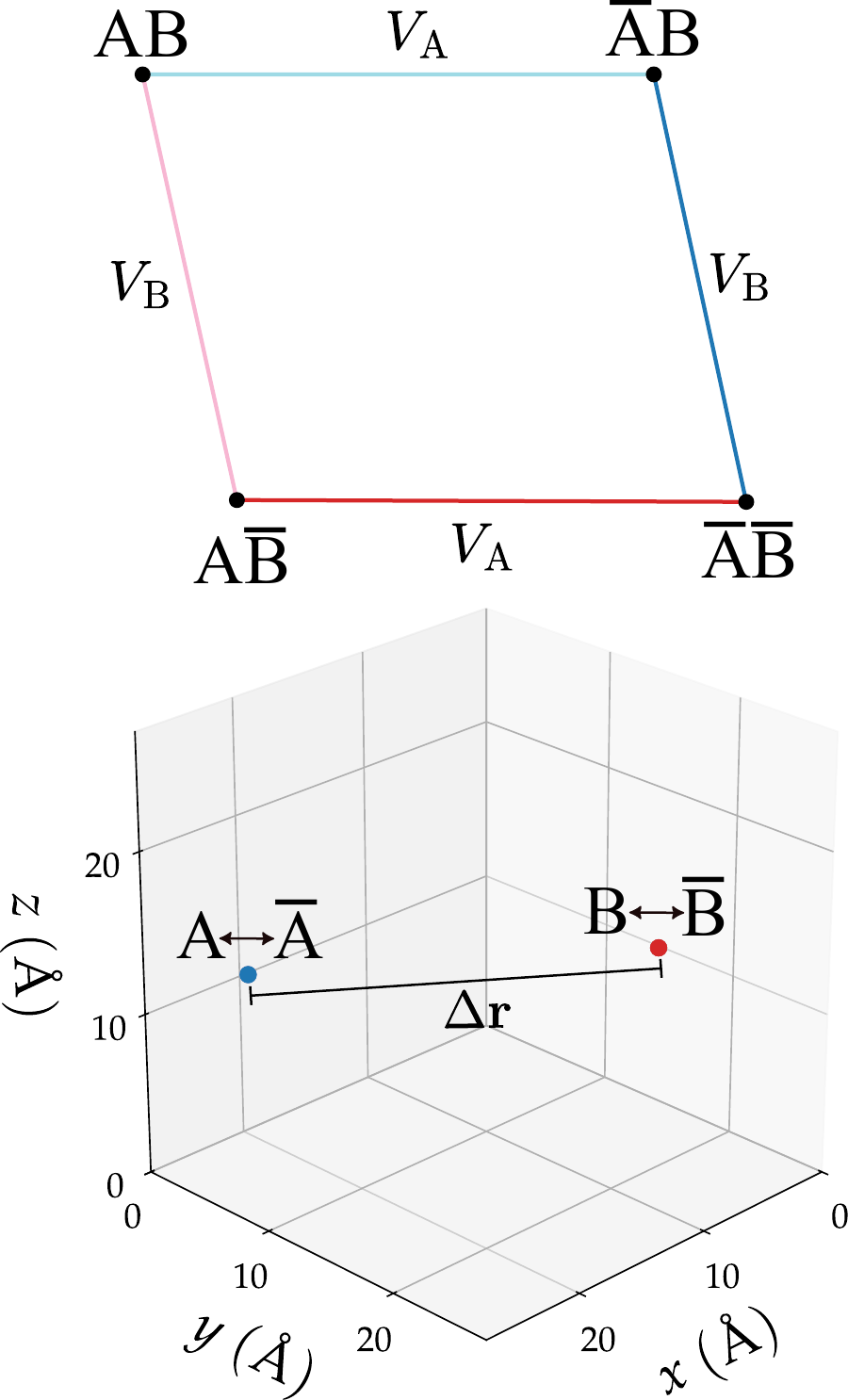}
    \caption{Two-level system like structure in a network formed by \textit{a-}Si. Top: Network formed by the four global states $\rm \{AB\}$ with barriers $V_{\rm A}$ and $V_{\rm B}$. Bottom: average position of the active atoms involved in two transitions in \textit{a-}Si. Since each transition has a beginning and end point, there are eight points in this figure (two of each color); however only two can be seen due to overlapping positions. The dark blue and pink overlap at $(x,y,z)\sim (20,0,10)$ \AA~while the light blue and red overlap at $(x,y,z)\sim (0,15,10)$ \AA. This implies that the overlapping colors may be the same TLS transitions which we label $A$ and $B$. The average distance between transition pair $\rm A\leftrightarrow\overline{A}$ and $\rm B\leftrightarrow\overline{B}$ is $\Delta\boldsymbol{r}$. Colors are matched between the top and bottom sub figures.}
	\label{fig: TLS Network}
\end{figure}

In order to provide a more precise comparison and quantification of the deviation from the TLS model and degree of coupling between distinct transitions, we must first understand the TLS model in the context of connected networks and the global energy landscape. A good example is given by a pair of transitions present in our \textit{a-}Si sample that behave like two independent TLS shown in Fig.~\ref{fig: TLS Network}. For two independent TLS that each transition between two configurations, namely $\rm A\leftrightarrow\overline{A}$ and $\rm B\leftrightarrow\overline{B}$, the global state-space has four possibilities $\rm \{AB , \overline{A}B, \overline{A}\overline{B}, A\overline{B}\}$. These states are connected in a symmetric four-state loop with energy barrier $V_{\rm A}$ for transitions involving $\rm A \leftrightarrow\overline{A}$ and $V_{\rm B}$ for $\rm B\leftrightarrow\overline{B}$. In this example, the distance $\Delta \boldsymbol{r}$ is relatively large, so the assumption of independence may be reasonable.

\begin{figure}
	\includegraphics[width=0.8\linewidth]{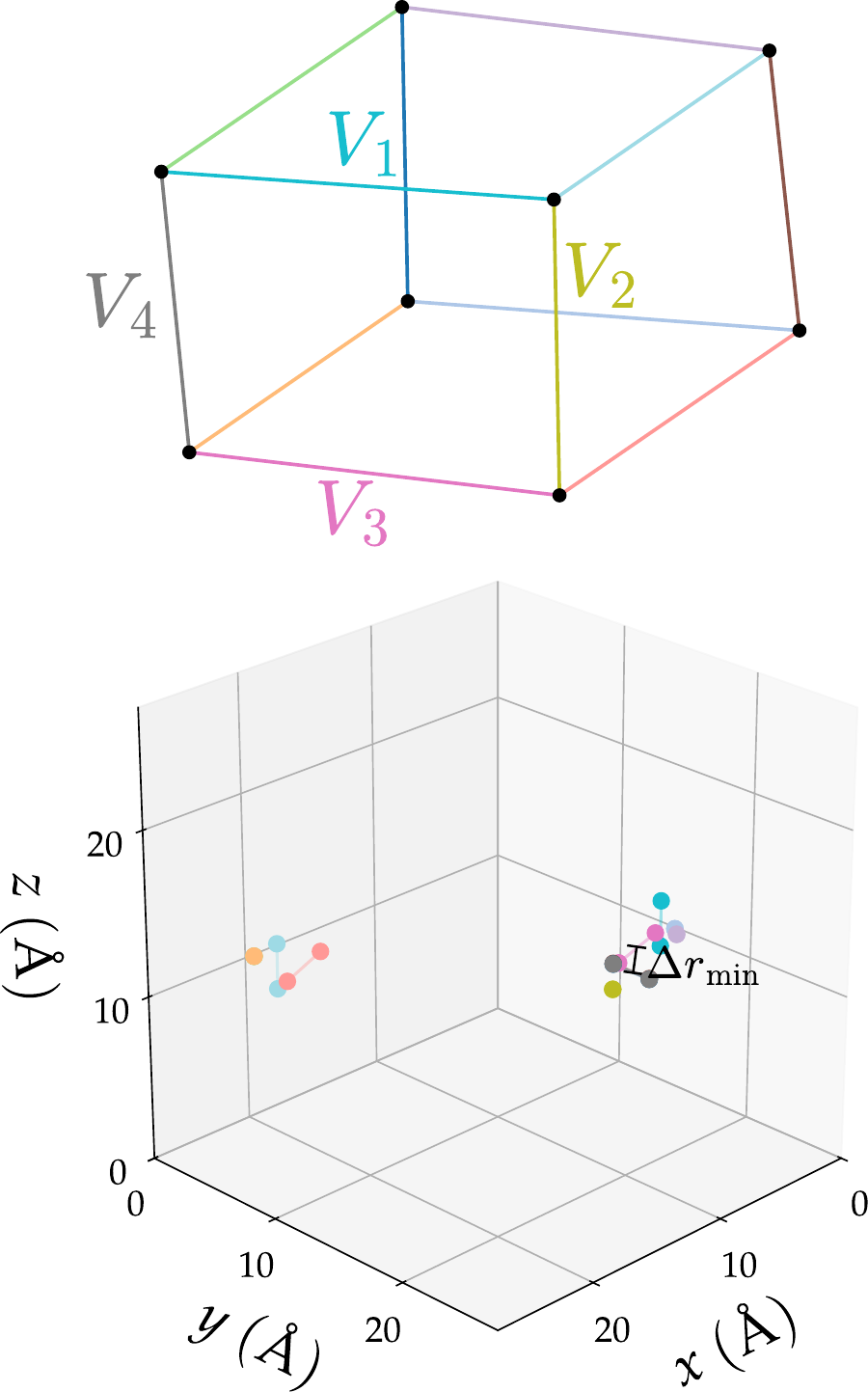}
    \caption{Cubic network structure present in the full network formed by a sample of \textit{a-}Si. Top: cubic network of transitions. One four-state cycle is labeled with energy barriers. For non-interacting TLS, both the even and odd edges must have equal energy barriers ($V_{1} = V_{3}$, $V_{2} = V_{4}$) and magnitude of asymmetries. Bottom: average position of the active atoms involved in the 12 transitions that make up the cubic network edges. We define $\Delta\boldsymbol{r}_{\rm min}$ as the shortest spatial distance between the average transition location of mismatched (even vs. odd) edges around each 4-state cycle. For the front face of the cube this corresponds to the pink (3) and grey (4) edges.}
	\label{fig: Cubic Network}
\end{figure}

We can now consider one higher level of complexity. Adding a third independent TLS $\rm C\leftrightarrow\overline{C}$  extends the global state-space to three dimensions, with states the permutations of $\rm \{A,B,C\}$ and their conjugate (overbar) states, which must form a cubic network (Fig.~\ref{fig: Cubic Network} top). Interestingly, we do observe cubic structures in our samples; however, they do not behave as independent TLS. While three independent TLS have three distinct real-space transition locations, in Fig.~\ref{fig: Cubic Network} (bottom) we find nine distinct real-space transition locations. In addition, several of the transitions take place in close proximity, implying that they may be strongly interacting. 

For any four-state closed loop, an independent TLS model predicts symmetric energy barriers and minima. In the context of Fig.~\ref{fig: TLS Network}, the odd (red) and even (blue) edges must have equal barriers and energy differences $V_{1} = V_{3},~|\Delta_{1}|=|\Delta_{3}|$ and $V_{2} = V_{4},~|\Delta_{2}|=|\Delta_{4}|$. We quantify the deviation from the TLS model by the largest difference around the loop $\Delta E^{\rm max} \equiv \max\{|V_{1}-V_{3}|,|V_{2}-V_{4}|,||\Delta_{1}|-|\Delta_{3}||,||\Delta_{2}|-|\Delta_{4}||\}$. The distance between transitions around a closed loop is defined to be the minimum root-mean-squared distance between the mean positions of active atoms involved in the odd and even edges $\Delta\boldsymbol{r}_{\rm min}^2 \equiv \min\{(\boldsymbol{r}_{1}-\boldsymbol{r}_{2})^2,(\boldsymbol{r}_{1}-\boldsymbol{r}_{4})^2,(\boldsymbol{r}_{3}-\boldsymbol{r}_{2})^2,(\boldsymbol{r}_{3}-\boldsymbol{r}_{4})^2\}$. For the labeled loop, this corresponds to the distance between the average location of the transition along the pink ($V_{3}$) and grey ($V_{4}$) edges (transitions).

\begin{figure}
	\includegraphics[width=\linewidth]{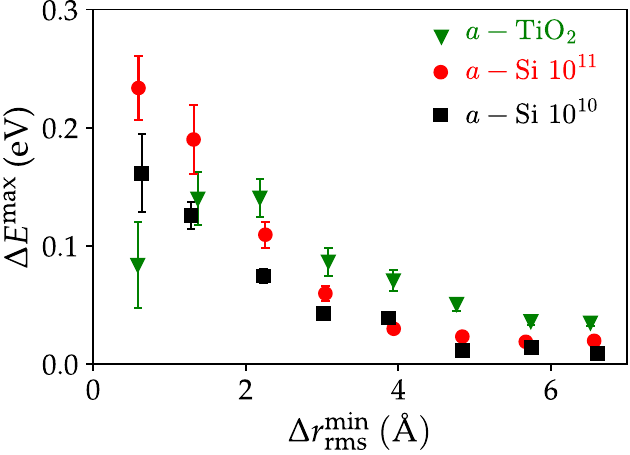}
    \caption{The maximum energy deviation from the TLS model $\Delta E^{\rm max}$ against the minimum root-mean-squared distance between the average transition locations $\Delta r^{\rm min}_{\rm rms}$ around all four-state cycles. Values are averaged over 10 samples of \textit{a-}Si ($10^{10}$K/s black squares, $10^{11}$K/s red circles) and \textit{a-}TiO$_{2}$ (green triangles) with error bars representing the standard error of the mean. A data point in our \textit{a-}Si sample at a quenchrate of $10^{10}$K/s with an energy barrier $\sim 4$eV and a $\Delta E^{\rm max}\sim 4$eV at $\Delta r^{\rm min}_{\rm rms}\sim 1$\AA~has been excluded from the average.}
	\label{fig: Deviation from TLS}
\end{figure}

At short range, we expect strong interactions between transitions. Indeed, this is reflected in Fig.~\ref{fig: Deviation from TLS}. Measuring the maximum deviation from a symmetric TLS for every four-state loop within our samples against their separation, we find significant changes in energy for transitions in close proximity. This, combined with the high propensity for transitions to cluster, implies that even when only accounting for the portions of the sample that act most like a TLS (four-state loops), we observe significant deviations from the TLS model.

\section{Quench rate}\label{sect: Quench}

 Now that we have established the importance of a network description, we turn to the implications of the network structure for the mechanical loss. The most consistent means of reducing mechanical loss is through annealing~\cite{khomenko2020,luckabauer2019,khadka2023,prasai2021}. It is well documented experimentally that better relaxed samples tend to have lower mechanical loss. We begin to explore this phenomenon by varying the quench rate in our sample preparation procedure. Unfortunately, with the quench rates accessible to our numerical simulations, we do not observe significant changes to our samples of \textit{a-}TiO$_{2}$ when decreasing the quench rate from $10^{11}$ to $10^{10}$K/s. Therefore, we will focus on \textit{a-}Si.

\begin{figure}
	\includegraphics[width=\linewidth]{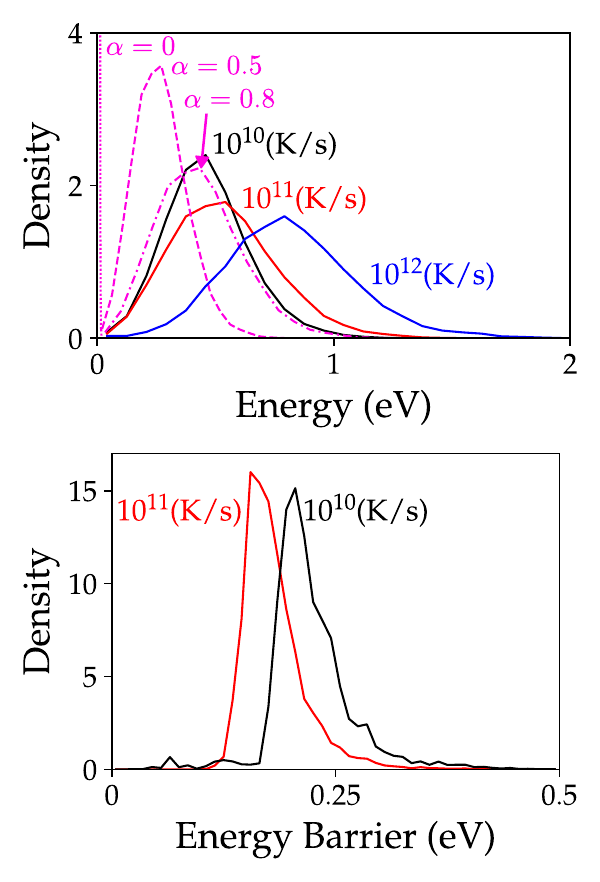}
    \caption{Summary of the impact of quench rate on the energy distribution of the network. Top: The distribution of inherent structure energies is shown for a fast quench of $10^{12}$K/s (blue), and two moderate quench rates of $10^{11}$ (red) and $10^{10}$K/s (black). Artificial annealing parameter $\alpha$ (purple) shifts the distribution towards a delta function as $\alpha\to 0$. Bottom: Average energy barrier for two moderate quench rates of $10^{11}$ and $10^{10}$K/s.}
	\label{fig: Energy and Barrier}
\end{figure}

\begin{figure}
	\includegraphics[width=\linewidth]{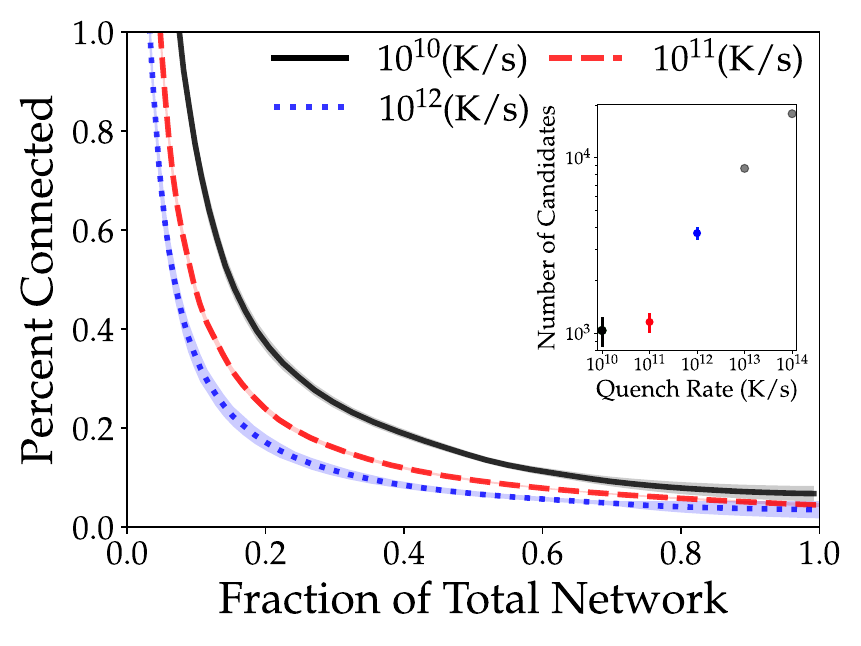}
    \caption{Fraction of the network that is connected (number of connections / number of possible connections) as a function of the fraction of the network that is sub-sampled at quenchrates of $10^{10}$ (black solid), $10^{11}$ (red dashed), and $10^{12}$K/s (blue dotted). The shaded regions denote the standard error of the mean. The inset shows the number of candidate transitions that could form part of the network as a function of the quenchrate.}
	\label{fig: Connectivity}
\end{figure}

Comparing quench rates from $10^{12}$ to $10^{10}$K/s for \textit{a-}Si, we observe significant changes to the energy landscape (Fig.~\ref{fig: Energy and Barrier}). As the quench rate is lowered, the system tends to form a more relaxed glass, not only by lowering the average energy of inherent states, but importantly by narrowing the distribution of inherent structure energies. In addition, the magnitude of the energy barriers increase on average; however, some small energy barrier are still present. This is accompanied by a monotonic increase in network connectivity as the quench rate is decreased (Fig.~\ref{fig: Connectivity}). Overall, we observe a systematic trend towards a more connected, flatter (equal energy minima) energy landscape with higher barriers at slower quench rates.

In our previous article \cite{Blaber2026}, we proposed two mechanisms that could decrease room temperature mechanical loss: high network connectivity, and a sharp distribution of energy minima. Neither of these can be described by the TLS model. To isolate and mimic the impact of annealing on the energy landscape, we artificially tighten the energy landscape by introducing an artificial annealing parameter $\alpha$, which changes the energy landscape without impacting the average barrier height:
\begin{align}
    E_i^{\alpha} &= \alpha E_i \nonumber \\
    E_j^{\alpha} &= \alpha E_j \nonumber \\
    V_{ij}^{\alpha} &= V^{\rm avg}_{ij} + \frac{1}{2}(E_i^{\alpha}+E_j^{\alpha}) \ ,
\end{align}
for $V_{ij}^{\rm avg} \equiv V_{ij} -(E_i+E_j)/2$.

Applying this to \textit{a-}Si at a quench rate of $10^{11}$K/s, we find that $\alpha = 0.8$ yields a similar distribution of energy minima as a quench rate of $10^{10}$ (Fig.~\ref{fig: Energy and Barrier}). As $\alpha\to 0$, the distribution approaches a delta function $P_{\alpha=0}(E_{j})=\delta(E_j - 0)$, with all inherent structures having equal energy minima.

\begin{figure}
	\includegraphics[width=\linewidth]{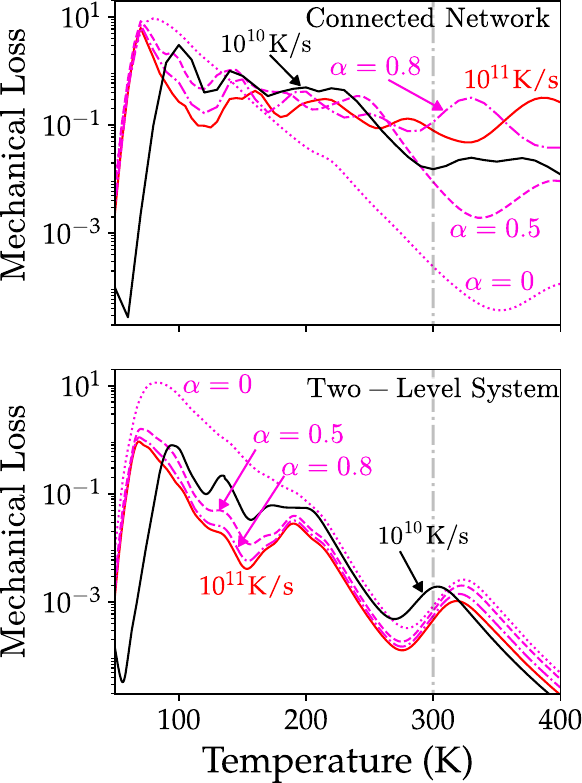}
    \caption{Impact of the energy landscape on the mechanical loss of \textit{a-}Si in the connected network (top) and TLS (bottom) models. Mechanical loss is calculated at $1000$Hz and is averaged over 10 samples each of \textit{a-}Si at a quenchrate of $10^{11}$K/s (red) and $10^{10}$K/s (black). As the parameter $\alpha$ is varied (purple), the energy landscape is monotonically and artificially shifted from the original distribution when $\alpha = 1$ to a delta-function distribution with all energy minima having $E = 0$ when $\alpha = 0$. For every transition, the average energy barrier is held fixed and is independent of $\alpha$.}
	\label{fig: Mechanical Loss}
\end{figure}

Comparing the mechanical loss between $10^{11}$ and $10^{10}$K/s at $300$K and $1000$Hz, we observe a significant decrease with slower quench rate  (Fig.~\ref{fig: Mechanical Loss}). This is in contrast to the TLS model, which for our data actually predicts a slight increase in room-temperature mechanical loss at the lower quench rate despite the reduced average number of transitions, due to the shift in the distribution of energy barriers and the reduction in energy asymmetry from the tightened energy distribution shown in Fig.~\ref{fig: Energy and Barrier}. Both models predict an increase in mechanical loss at moderate temperatures, with a higher temperature peak and subsequent drop off in mechanical loss at $10^{10}$K/s. This peak directly reflects the shift in the peak of the barrier height distribution in Fig.~\ref{fig: Energy and Barrier}. Overall, at low temperatures mechanical loss predicted by the connected network and TLS models appear qualitatively similar within this classical description.

To isolate the effect of the distribution of energy minima, we turn to our artificial annealing parameter $\alpha$ (Fig.~\ref{fig: Mechanical Loss}). In the TLS model, the impact is easy to understand: any tightening of the distribution of energy minima decreases asymmetry between TLS and monotonically increases mechanical loss. However, in the more precise description of the connected network model, the effect is reversed. As artificial annealing is strengthened ($\alpha\to 0$), room temperature (high T) mechanical loss decreases. In this case, the TLS model disagrees qualitatively with experimental trends.

\begin{figure}
	\includegraphics[width=\linewidth]{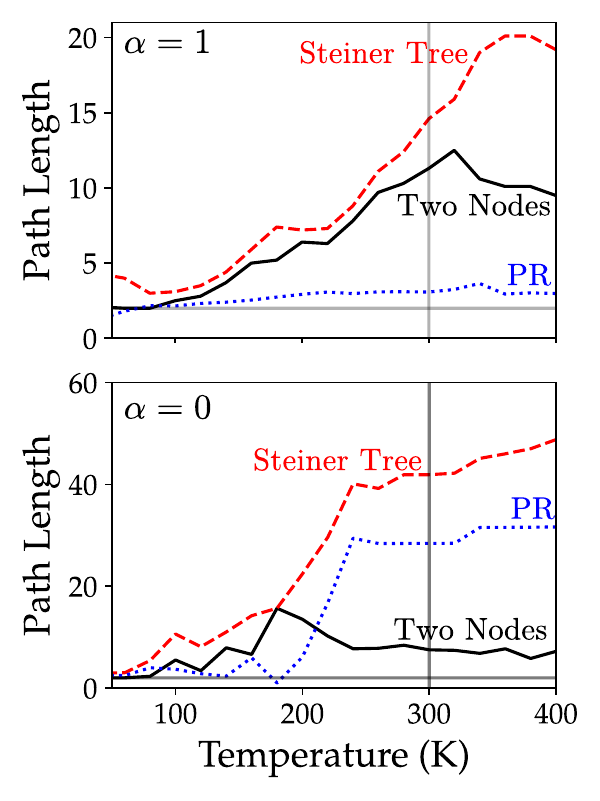}
    \caption{Average network path length of the eigenvector with the most significant contribution to mechanical loss as a function of temperature. All quantities are averaged over 10 samples of \textit{a-}Si at a quenchrate of $10^{11}$ and artificial annealing parameter $\alpha = 1$ (top) and $\alpha = 0$ (bottom). The path length is estimated by the number of nodes in the shortest path through the network between the two largest components of the eigenvector (two nodes, black solid curve), the eigenvector participation ration (PR, blue dotted), and the number of nodes in the Steiner tree (red dashed) connecting the $\max\{\lceil {\rm PR}\rceil,2\}$ largest components of the eigenvector. Grey lines mark a path length of two and a temperature of $300$K. We have excluded one data point from $\alpha = 0$ averages at $180$K with a Steiner tree length of 479 nodes, PR of 474, and two node length of 40.}
	\label{fig: path length}
\end{figure}

Since the reduction in mechanical loss upon tightening the distribution of energy minima is a purely network phenomenon, we turn to the network properties to better understand this observation. Our goal will be to better understand the energetics of the path through network space that the system takes.
At each temperature, we find the eigenvector of the transition rate matrix that has the largest contribution to mechanical loss and compute the participation ratio
\begin{align}
    [{\rm PR}]_{j} \equiv \frac{\left(\sum_i M_{ij}^2\right)^2}{\sum_i M_{ij}^4} \ .
\end{align}
The participation ratio quantifies how many elements of the eigenvector are significant. For a participation ratio of two, only two elements significantly contribute to the eigenmode. If this is the case, then a reasonable guess for the transition path taken between these two nodes is given by the shortest path connecting them within the network. The average length of this path is labeled as `Two Nodes' in Fig.~\ref{fig: path length}; however, the participation ratio is generally larger than two. For participation ratios larger than two, the path is better represented by the Steiner tree, the smallest network that connects all the nodes that contribute significantly to the eigenvector (the $\max\{2,\lceil PR\rceil$\} nodes), with length given by the number of nodes in the Steiner tree network.

In general, we find that the paths taken through network space by the dominant eigenmodes increase in length with increasing temperature (Fig.~\ref{fig: path length}). If we equate TLS-like behavior with a path length of two, then as temperature is increased the system becomes increasingly dissimilar to isolated TLSs. Furthermore, we observe a marked increase in path length when the energy distribution is artificially annealed to a delta-function distribution.

\begin{figure}
	\includegraphics[width=\linewidth]{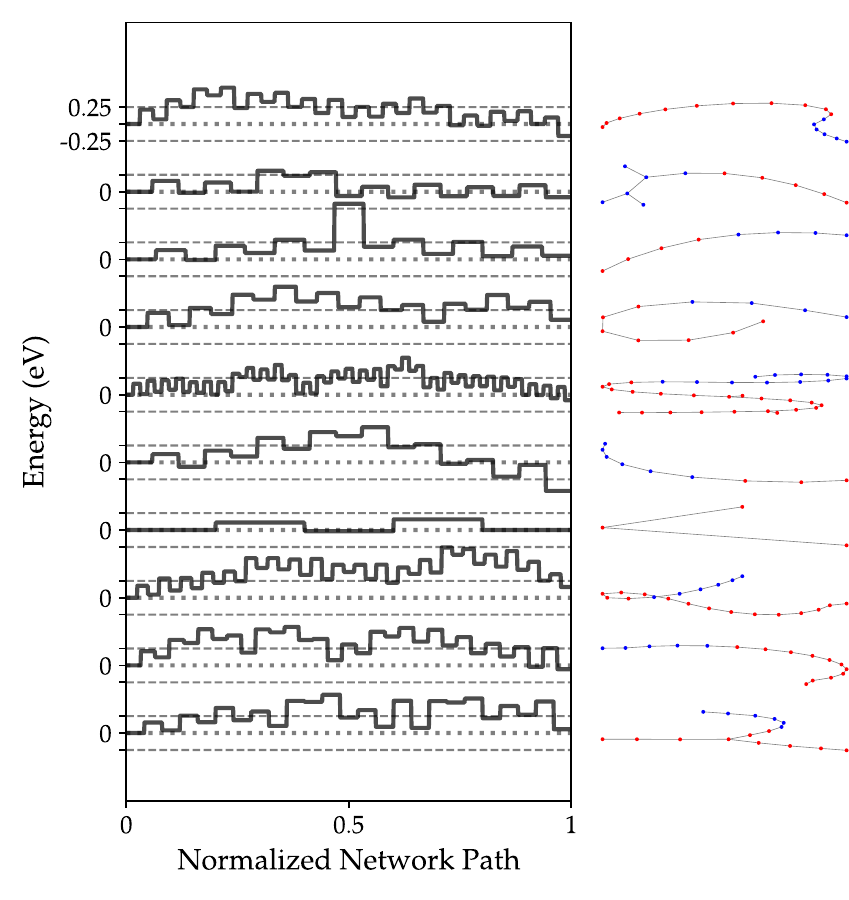}
    \caption{Energy landscape (left) along the backbone (longest linear path) of the Steiner tree (right) connecting the $\max\{\lceil {\rm PR}\rceil,2\}$ largest components of the dominant eigenvector at $300$K and $1000$Hz. Ten independent samples of \textit{a-}Si at a quenchrate of $10^{11}$K/s are shown offset in energy with no artificial annealing $\alpha = 1$. The color (red/blue) of the nodes in the network denote the relative sign of the corresponding element of the eigenvector.}
	\label{fig: path energy alpha 1}
\end{figure}

A major benefit of this form of analysis is that it yields direct access to the one-dimensional energy landscape connecting the dominant elements of the dominant eigenmode. In Fig.~\ref{fig: path energy alpha 1} we plot the one-dimensional path taken along the backbone (longest one-dimensional chain) of the Steiner tree of the dominant elements of the eigenmode with the largest contribution to mechanical loss at $300$K and $1000$Hz.

\begin{figure}
	\includegraphics[width=\linewidth]{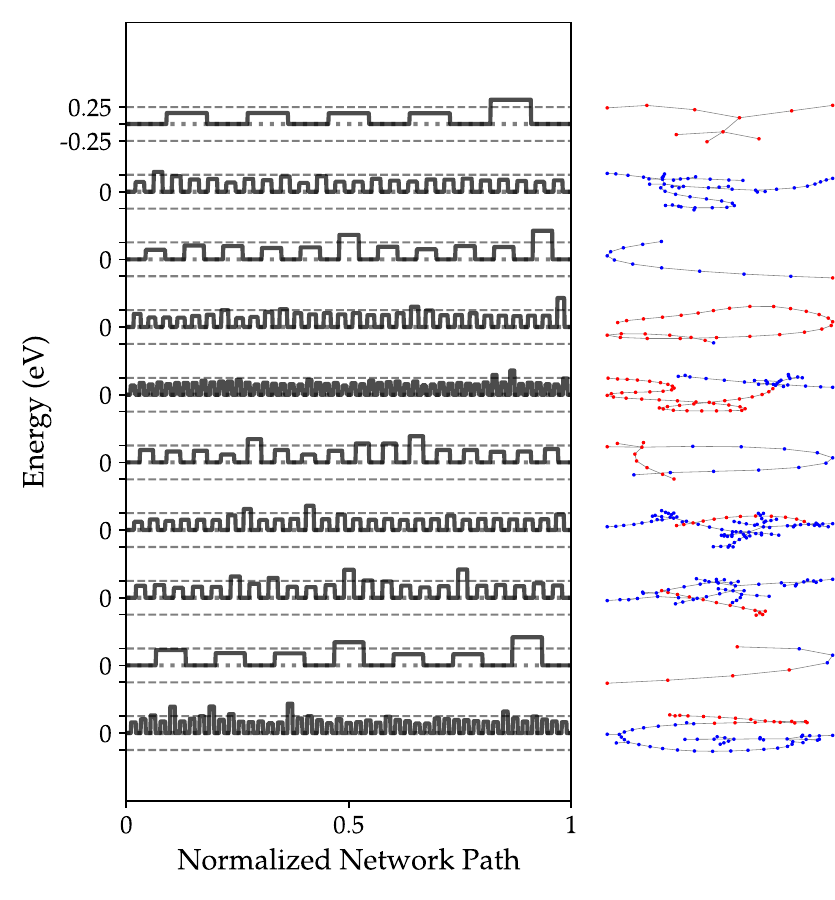}
    \caption{Energy landscape (left) along the backbone (longest linear path) of the Steiner tree (right) connecting the $\max\{\lceil {\rm PR}\rceil,2\}$ largest components of the dominant eigenvector at $300$K and $1000$Hz. Ten independent samples of \textit{a-}Si at a quenchrate of $10^{11}$K/s are shown offset in energy with maximal artificial annealing $\alpha = 0$. The color (red/blue) of the nodes in the network denote the relative sign of the corresponding element of the eigenvector.}
	\label{fig: path energy alpha 0}
\end{figure}

Notably, the paths appear to have a mountain structure, with higher energy minima toward the middle of the path and lower minima at the edges. This type of structure was analytically and numerically shown to give high temperature (low frequency) mechanical loss in our previous article~\cite{Blaber2026}. This confirms that this structure may be the root cause of high temperature mechanical loss within our samples. Indeed, when $\alpha\to 0$ the paths remain long but no longer have a mountainous structure (Fig.~\ref{fig: path energy alpha 0}), resulting in the observed reduction in room temperature mechanical loss.

There is another dissipation mode present in Fig.~\ref{fig: path energy alpha 1} which does not appear to have the mountain structure and only has three nodes present. Accounting for the sign of the eigenvector, all the nodes present have the same sign. In this case, the eigenmode corresponds to three dominant nodes interacting with a broad and diffuse background: probability flows between these three nodes and a large number of background nodes. In order for this eigenmode to dominate at high-temperatures it must be a slow relaxation mode. This could be formed by either a topological bottleneck between the diffuse background and the dominant nodes or an intermediate region of low probability (high energy minima) separating the dominant nodes and the diffuse background. Since this dissipation mode is no longer present when the energy landscape is flattened ($\alpha = 0$), the latter is likely the case. This represents a novel, slow relaxation pathway that nevertheless acts and appears similar to the mountain landscape. Although this is only seen in one sample of \textit{a-}Si, this dissipation mechanisms accounts for the room temperature loss in nine out of ten samples of \textit{a-}TiO$_{2}$ (Appendix.~\ref{app: App 1}).

\section{Discussion}\label{sect: Discussion}

In this article, we have provided more detail and context for the recently introduced network description of mechanical loss in amorphous solids~\cite{Blaber2026}. In section~\ref{sect: Spatial}, we analyzed in detail the parts of network structure that appear most similar to TLS (four-state cycles), and found that they deviate significantly from non-interacting TLSs by up to $\sim 0.25$eV within an interaction range of $\sim4$\AA. This range is similar to the scale of localized clustering of transitions that we observe with a radius of up to $\sim4$\AA~(Fig.~\ref{fig: Deviation from TLS}). The origin of this (relatively short) length scale is unclear at the moment, but might be linked to dynamical heterogeneities observed in glass forming liquids. This highlights the importance of a network description. Even accounting for structures most similar to TLSs, we observe significant interactions between transitions, especially at short range. Since the transitions have a tendency to be locally clustered, there is a high probability for transitions to be within the effective interaction range.

To improve our estimates of local clustering and interaction range, it will be necessary to simulate larger systems, better annealed samples, and with improved molecular potentials. Larger systems would allow for more accurate measures of interactions between TLS which could give more insight into the nature of the interactions, ranging from direct steric repulsion to long range elastic or electrostatic interactions. Simulations of well annealed and ultra-stable glasses would be interesting since the defect and TLS density is expected to decrease significantly; however, if the transitions remain locally clustered they may be strongly interacting despite a relatively low density. 

A recent study found differences in the nature and number of defects depending on the interatomic potential employed~\cite{girard2025}. To account for the impact of the interatomic potential, we have studied two different materials with distinct types of molecular potentials, an elemental network glass and a metal oxide with two elements and electrostatic interactions. As in our previous article, we aim to explore general trends rather than material specific properties~\cite{Blaber2026}. We expect our main results to be robust to model details: interacting and locally clustered transition, tightening of the energy landscape for better relaxed samples, and a reduction in high temperature mechanical loss with a tightening of the energy landscape. However, details such as the length scale of interacting transitions, distribution of energy barriers, distribution of energy minima, and the overall number and density of transitions may be strongly dependent on the details of the material and interatomic potential.

Assuming a simple form of exponentially decaying radial distribution function $g(r) = 1+A\exp\{-r/\xi\}$, the number of transitions within a radius $R$ is $N(R) = 4\pi n \int_{0}^{R}\md r~r^2g(r)$ for average density $n = N_{\rm transitions}/V$. Assuming $A\approx 10^2$ and $\xi \approx 1$ (Fig.~\ref{fig: Deviation from TLS}), finding two transitions within a range of $N(8{\rm \AA}) = 2$ then requires an average transition density of $n\sim 0.3{\rm nm}^{-3}$. For amorphous silicon with a density of $2.28{\rm g/cm}^{3}$, this corresponds to $\sim 0.6$\% atomic defect density, comparable to coordination defect densities for more sophisticated molecular potentials~\cite{girard2025}. Interestingly, experimental estimates of TLS densities in \textit{a-}Si vary over four orders of magnitude, from $160 {\rm eV}^{-1}{\rm nm}^{-3}$ to $0.016{\rm eV}^{-1}{\rm nm}^{-3}$~\cite{queen2015}. Assuming an energy range of $\sim 1$eV, only the low defect samples are below our $n\sim 0.3{\rm nm}^{-3}$ heuristic for interacting transitions. This is only meant as an order of magnitude estimate; the specific value depends strongly on the interaction length $R$ and the correlation length $\xi$. It should also be noted that experimental estimates of the TLS density themselves rely on the TLS model.

Towards assessing the impact of annealing, we have studied in section \ref{sect: Quench} the impact of a slower quench rate, which leads to better annealed samples. We observe that a slower quench rate not only lowers the average energy of the samples, but also tightens the energy barrier distribution. To isolate and amplify the impact of changes to the energy landscape, we introduced an artificial annealing parameter and showed that a tightening of the energy landscape can eliminate high temperature mechanical loss (Fig.~\ref{fig: Mechanical Loss}). This is in contrast to the TLS model, which predicts the opposite effect, namely that a tightened energy distribution increases mechanical loss. We explored in detail the mechanism behind this reduction in high temperature mechanical loss from a network and energy landscape perspective (Figs.~\ref{fig: path energy alpha 1} and \ref{fig: path energy alpha 0}).

It will be interesting to explore the implications for low temperature mechanical loss. Superficially, it appears that our estimates from the connected network and TLS models are qualitatively similar at low temperature; however, we have not accounted for corrections to the transition rates due to tunneling effects which become relevant below 10K. In addition, due to the high precision and fidelity of dielectric loss measurements, it is already necessary and standard practice to employ interacting TLS models to properly model dielectric loss from the amorphous solids in superconducting qubits~\cite{clare1989interacting,carruzzo1994nonequilibrium,martinis2005decoherence,klimov2018fluctuations,muller2019,bejanin2021interacting}. The network model described here can be thought of as an interacting TLS model with one fewer level of assumptions. Therefore, it would be interesting to see what implications it may have for low temperature dielectric and mechanical loss. Direct spectroscopic control and study of dielectric loss in amorphous coatings offer a very promising path towards a precise understanding of dissipation mechanisms in amorphous solids~\cite{wang2025evidence,wang2025Spectroscopy}.

\section*{Acknowledgments}
This research was supported in part by a Natural Sciences and Engineering Research Council of Canada (NSERC) Canada Postdoctoral Research Award (SB), the Canada First Research Excellence Fund, Quantum Materials and Future Technologies Program, the New Frontiers in Research Fund (Exploration stream), and the Discovery Grant program of the Natural Sciences and Engineering Research Council of Canada (JR). Computational resources and services were provided by Advanced Research Computing at the University of British Columbia and the Digital Research Alliance of Canada (\url{alliancecan.ca}).

\setcounter{section}{0}

%%%%%%%%%% Merge with supplemental materials %%%%%%%%%%
\onecolumngrid
\clearpage
\begin{center}
	\textbf{\large Supplemental Material for ``Connected Network Model for the Mechanical Loss of Amorphous Materials''}
\end{center}
%%%%%%%%%% Merge with supplemental materials %%%%%%%%%%
%%%%%%%%%% Prefix a "S" to all equations, figures, tables and reset the counter %%%%%%%%%%
\setcounter{equation}{0}
\setcounter{figure}{0}
\setcounter{table}{0}
\setcounter{page}{1}
\makeatletter
\renewcommand{\theequation}{S\arabic{equation}}
\renewcommand{\thefigure}{S\arabic{figure}}

\section{\textit{a-}TiO$_{2}$}\label{app: App 1}

\begin{figure}[h!]
	\includegraphics[width=0.5\linewidth]{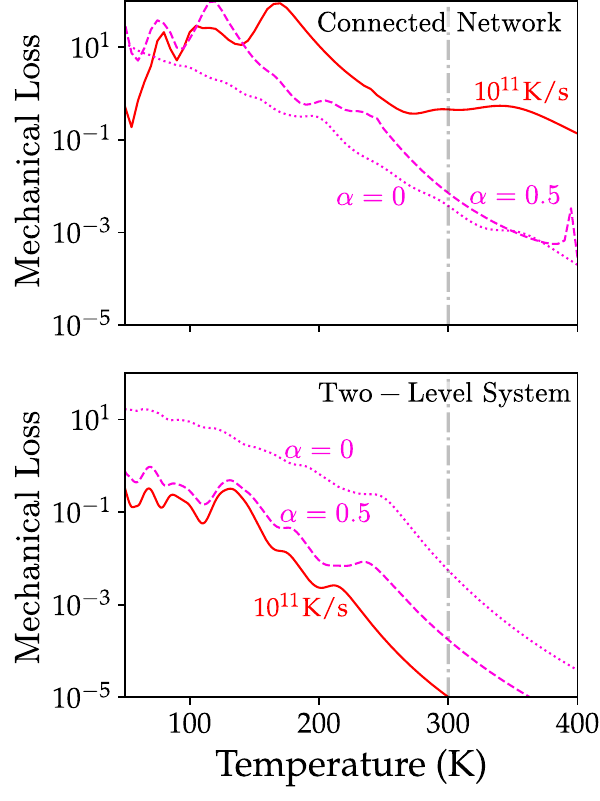}
    \caption{Impact of the energy landscape on the mechanical loss of \textit{a-}TiO$_{2}$ in the connected network (top) and TLS (bottom) models. Mechanical loss calculated at $1000$Hz and is averaged over 10 samples at a quenchrate of $10^{11}$K/s. As the parameter $\alpha$ is varied, the energy landscape is monotonically and artificially shifted from the original distribution when $\alpha = 1$ to a delta-function distribution with all energy minima having $E = 0$ when $\alpha = 0$. For every transition, the average energy barrier is held fixed and is independent of $\alpha$.}
	\label{fig: Ti Mechanical Loss}
\end{figure}

In this section we provide an analysis analogous to the main text applied to \textit{a-}TiO$_{2}$. Applying artificial annealing through the parameter $\alpha$ in Fig.~\ref{fig: Ti Mechanical Loss}, we observe similar trends to \textit{a-}Si.  As $\alpha\rightarrow0$ the energy distributions tightens, leading to a decrease in high temperature mechanical loss in the connected network description and an increase in mechanical loss from the TLS model (Fig.~\ref{fig: Ti Mechanical Loss}). Our barrier distribution and TLS model predictions for mechanical loss in \textit{a}-TiO$_{2}$ are qualitatively similar to Ref.~\cite{trinastic2016}, with a low temperature peak followed by a significant drop in mechanical loss at high temperatures.

\begin{figure}
	\includegraphics[width=0.5\linewidth]{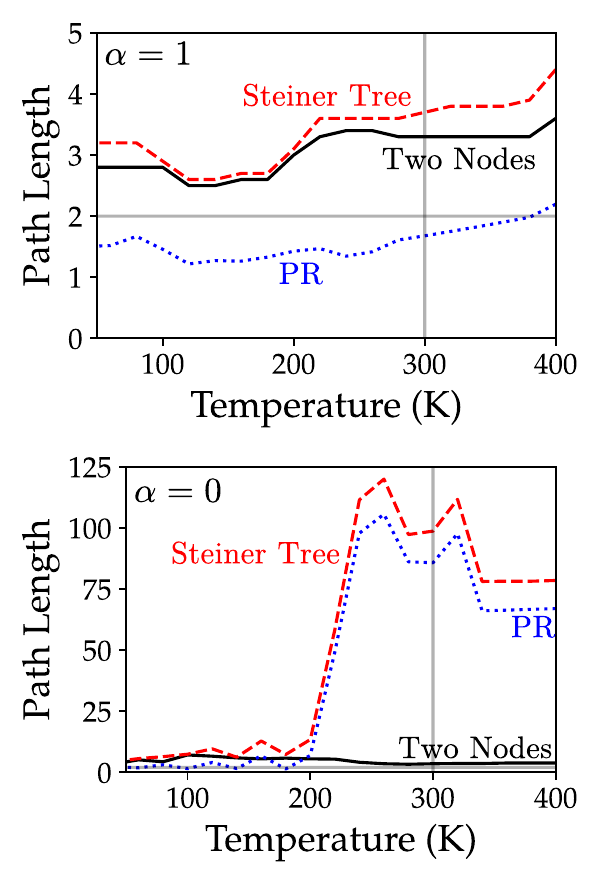}
    \caption{Network path length of the eigenvector with the most significant contribution to mechanical loss as a function of temperature. All quantities are averaged over 10 samples of \textit{a-}TiO$_{2}$ at a quenchrate of $10^{11}$ and artificial annealing parameter $\alpha = 1$ (top) and $\alpha = 0$ (bottom). The path length is estimated by the number of nodes in the shortest path through the network between the two largest components of the eigenvector (two nodes, black solid curve), the eigenvector participation ration (PR, blue dotted), and the number of nodes in the Steiner tree (red dashed) connecting the $\max\{\lceil {\rm PR}\rceil,2\}$ largest components of the eigenvector. Grey lines mark a path length of two and a temperature of $300$K. }
	\label{fig: Ti path length}
\end{figure}

\begin{figure}
	\includegraphics[width=0.5\linewidth]{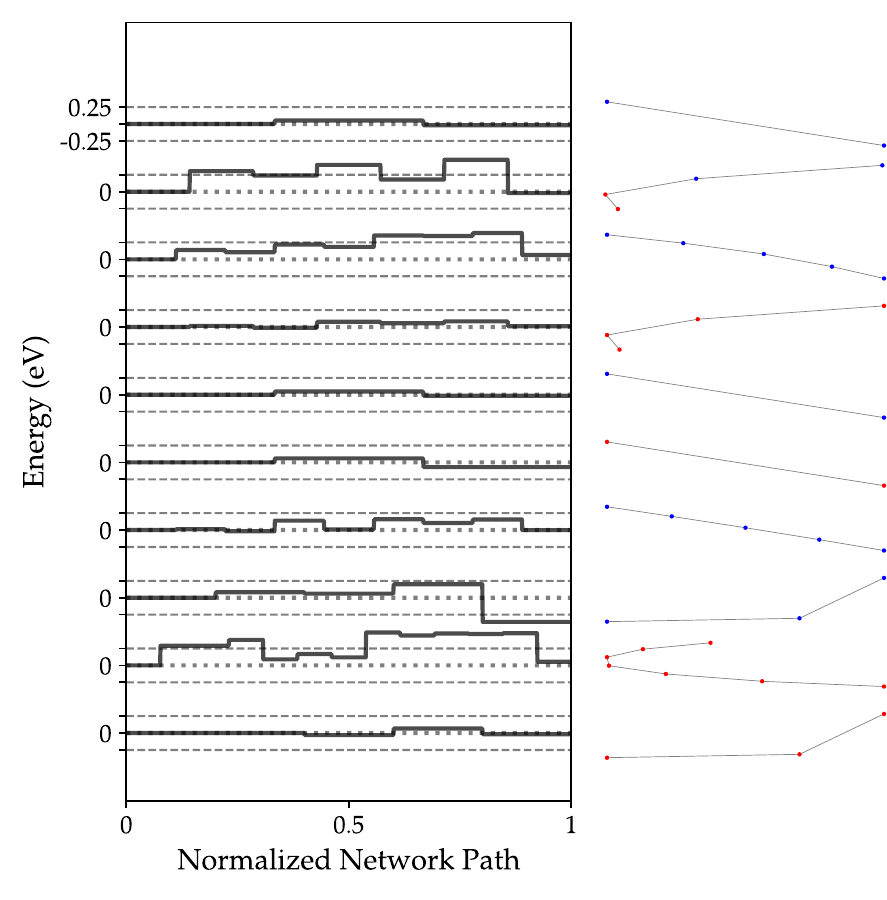}
    \caption{Energy landscape (left) along the backbone (longest linear path) of the Steiner tree (right) connecting the $\max\{\lceil {\rm PR}\rceil,2\}$ largest components of the dominant eigenvector at $300$K and $1000$Hz. Ten independent samples of \textit{a-}TiO$_{2}$ at a quenchrate of $10^{11}$K/s are shown offset in energy with no artificial annealing $\alpha = 1$. The color (red/blue) of the nodes in the network denote the relative sign of the corresponding element of the eigenvector.}
	\label{fig: Ti path energy alpha 1}
\end{figure}

We compute the participation ratio of the dominant eigenmodes and their path lengths as described in the main text (Fig.~\ref{fig: path length}), now for \textit{a-}TiO$_{2}$ in Fig.~\ref{fig: Ti path length}. For $\alpha=1$ (the unchanged raw data), we observe significant differences in the properties of the eigenmodes compared to \textit{a-}Si. Notably, the participation ratio often lies below two and the path lengths are significantly shorter than those found in \textit{a-}Si. This can be explained by considering the specific energy landscapes and networks formed by the dominant eigenmodes at $300$K shown in Fig.~\ref{fig: Ti path energy alpha 1}. All but one of the samples have a relatively small Steiner tree of largest eigenvector components that all have the same sign. This implies that the nodes shown do not tell the whole picture despite being dominant in magnitude. These eigenvectors correspond to a few dominant nodes exchanging probability with a large diffuse background of nodes, each of which has a small contribution to the total eigenvector individually. 

For an eigenvector with $m$ dominant components with magnitude $a$ and $N-m$ non-dominant components with magnitude $b/(N-m)$, the participation ratio is 
\begin{align}
{\rm PR} &= \frac{\left[m a^2+\frac{b^2}{N-m}\right]^2}{m a^4+\frac{b^4}{(N-m)^3}} \\
&\approx m+\frac{2b^2}{a^2N}+O\left(N^{-2}\right) \ .
\end{align}
If $m\sim 1$ and $N$ is large then the participation ratio is of order one, consistent with Fig.~\ref{fig: Ti path length}.

\begin{figure}
	\includegraphics[width=0.5\linewidth]{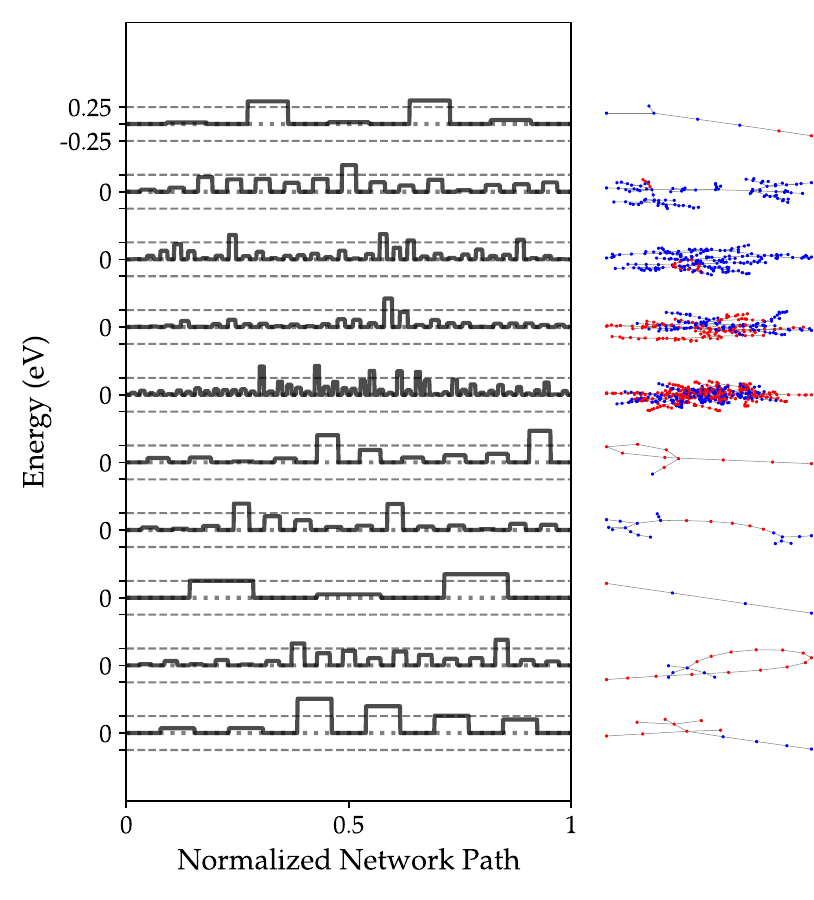}
    \caption{Energy landscape (left) along the backbone (longest linear path) of the Steiner tree (right) connecting the $\max\{\lceil {\rm PR}\rceil,2\}$ largest components of the dominant eigenvector at $300$K and $1000$Hz. Ten independent samples of \textit{a-}TiO$_{2}$ at a quenchrate of $10^{11}$K/s are shown offset in energy with maximal artificial annealing $\alpha = 0$. The color (red/blue) of the nodes in the network denote the relative sign of the corresponding element of the eigenvector.}
	\label{fig: Ti path energy alpha 0}
\end{figure}

When the samples are artificially annealed, high temperature dissipation decreases (Fig.~\ref{fig: Ti Mechanical Loss}) and the path length increases dramatically (Fig.~\ref{fig: Ti path length}). This is reflected in the paths taken by the dominant elements of the dominant eigenmodes at $300$K in Fig.~\ref{fig: Ti path energy alpha 0}. This implies that the structure of the energy landscape is what causes the eigenmodes observed for $\alpha=1$ to cause significant high temperature mechanical loss. In order for an eigenmode to dominate at $300$K with the energy barriers present in our samples, it must be a slow relaxation mode. There are two ways this concentrated nodes exchanging with a diffuse background can result in a relaxation mode much slower than the intrinsic relaxation time of the individual transitions: 1. a topological bottleneck where the pathway between the concentrated nodes and the diffuse background has very few pathways and 2. an energetic bottleneck caused by high energy (low probability) states separating the dominant nodes from the diffuse background. The former mechanism is topological in nature and should not depend strongly on the energy landscape, so the latter is the likely culprit for the high temperature mechanical loss found in our samples. This present a novel dissipation mechanism that is similar in structure to the slow mountain dissipation mode which is primarily observed in our \textit{a-}Si samples.

\end{document}